\documentclass[prb,superscriptaddress,twocolumn,showpacs]{revtex4-1}
\usepackage{amsmath}
\usepackage{amsfonts}
\usepackage{amssymb}
\usepackage{graphicx,graphics,color}

\newcommand{\beq}{\begin{equation}}
\newcommand{\eeq}{\end{equation}}
{
 \definecolor{BLACK}{gray}{0}
 \definecolor{WHITE}{gray}{1}
 \definecolor{RED}{rgb}{1,0,0}
 \definecolor{GREEN}{rgb}{0,1,0}
 \definecolor{BLUE}{rgb}{0,0,1}
 \definecolor{CYAN}{cmyk}{1,0,0,0}
 \definecolor{MAGENTA}{cmyk}{0,1,0,0}
 \definecolor{YELLOW}{cmyk}{0,0,1,0}
}

\begin{document}

\title{Reply to Comment on "Superconductivity at low density near a
ferroelectric quantum critical point: doped SrTiO$_{3}$" }
\author{Peter W\"{o}lfle}
\affiliation{Institute for Condensed Matter Theory and Institute for Nanotechnology, Karlsruhe Institute of Technology, 76021 Karlsruhe, Germany}

\author{Alexander V. Balatsky }
\affiliation{Nordita, Stockholm, SE 10691, Sweden}

\date{\today }

\begin{abstract}
In our paper (Wölfle and Balatsky, Phys. Rev. B 98, 104505 (2018)) 
we presented a microscopic theory of
superconductivity for doped SrTiO$_{3}$ by proposing two pairing mechanisms
acting simultaneously with relative strength depending on the closeness to
the ferroelectric quantum critical point. The first mechanism rests on the
dynamically screened Coulomb interaction, and the second assumed a coupling
to the soft transverse optical phonon. In their comment Ruhman and Lee 
point out an error in our estimate of the deformation
potential coupling to the soft mode. We agree that this type of coupling
cannot explain the gigantic isotope effect observed experimentally, so that
a different coupling mechanism needs to be found. As for the first pairing
mechanism, Ruhman and Lee maintain the view expressed in their paper 
(Ruhman and Lee, Phys. Rev. B 94, 224515 (2016)) that the energy range 
over which the usual longitudinal optical
phonon mediated interaction operates is limited by the Fermi energy. We
object to this view and in this reply present evidence that the cutoff
energy is much larger. In a weak coupling system such as SrTiO$_{3}$ the
cutoff is given by the energy beyond which quasiparticles cease to be well
defined.
\end{abstract}

\pacs{}
\maketitle

\affiliation{Institute for Condensed Matter Theory and Institute for
Nanotechnology, Karlsruhe Institute of Technology, 76021 Karlsruhe, Germany}

\affiliation{Nordita, Stockholm, SE 10691, Sweden}

\affiliation{Institute for Condensed Matter Theory and Institute for
Nanotechnology, Karlsruhe Institute of Technology, 76021 Karlsruhe, Germany}

\affiliation{Nordita, Stockholm, SE 10691, Sweden}

\section{Introduction}

In their Comment \cite{Ruhman19} Ruhman and Lee criticize our derivation of
the cutoff frequency in the gap equation, $\omega _{c}$, by arguing that at
energies beyond the Fermi energy $\epsilon _{F}$ an uncontrollable set of
additional contributions in perturbation theory would have to be taken into
account. We object to this statement as unfounded, as explained in detail
below. As for the second part of our paper \cite{Wolfle18}, devoted to the
explanation of the isotope effect, we agree with\ Ruhman and Lee \cite%
{Ruhman19}\ \ to have made an error in evaluating the deformation potential
coupling of carriers to the soft TO phonon mode. This coupling is indeed
negligible and the results for the transition temperatures $T_{c}$ obtained
by using that interaction are incorrect. In the following we present results
of a reevaluation of $T_{c}$ obtained by approximating the pair interaction
by the screened Coulomb interaction. We give the specifics below.

In their paper on superconductivity in lightly doped SrTiO$_{3}$ \cite%
{Ruhman16} Ruhman and Lee, (1) apply the usual method of separating pair
interaction into a longitudinal optical (LO) phonon mediated attractive part
and a repulsive Coulomb interaction part, which they treat by subtracting a $%
\mu ^{\ast }$ parameter of conventional magnitude to the pair coupling. This
conventional procedure requires the Fermi energy $\epsilon_{F}\gg\omega_{LO}$%
, the optical phonon frequency and thus is no longer valid once the Fermi
energy falls below the longitudinal phonon energy, as is the case for
densities $n< 3*10^{20}$cm$^{-3}$.  They further assume, (2) a frequency
cutoff in the gap equation of order Fermi energy $\epsilon _{F}$, without
justification. The resulting transition temperatures calculated by them are
orders of magnitude too low at low density \cite{Behnia13,Behnia14,Schooley64}. 
They conjecture that the doping
reduces the dielectric constant by a factor of $\approx 20$ even at the low
doping end. There is no experimental indication for such a drastic change of
the dielectric properties induced by light doping.

We believe to have resolved both difficulties, (1) and (2), in our work \cite%
{Wolfle18}, in which (1) we use the fully screened Coulomb interaction, thus
avoiding (a) the splitting off of the phonon mediated interaction and (b)
treating the remaining static Coulomb interaction only approximately. We (2)
show that the ionic screening is so strong as to keep the dimensionless
screened Coulomb interaction in the weak coupling regime not only for
energies less than the Fermi energy, but up to the LO phonon energy scale. A
detailed evaluation of the dimensionless coupling function is presented
below, demonstrating that the coupling is indeed much less than unity for
typical momenta and frequencies up to the phonon scale. The frequency cutoff 
$\omega _{c}$\ in the gap equation is provided by the energy beyond which
the electron self energy exceeds the energy itself. We present below a
calculation of the transition temperature again using a hard cutoff
approximation, for calculational convenience. The resulting transition
temperature as a function of doping is in good agreement with 
experiment \cite{Behnia13,Behnia14,Schooley64} .

As for the isotope effect \cite{vdMarel16} we do not yet have a convincing derivation of the
coupling of charge carriers to the soft TO phonon of the required strength,
ready to be reported here. We note, however, that additional support for the
existence of the strong isotope effect is coming from the observed gigantic
increase of $T_{c}$ by applying pressure and tensile strain, moving the
system closer to the QCP \cite{Sochnikov18}.

\section{Superconductivity mediated by the screened Coulomb interaction}

The superconducting state of an interacting Fermi system is characterized by
a gap parameter $\Delta (\mathbf{k,}i\omega _{n})$ obtainable from the gap
equation \cite{Eliashberg60}

\begin{equation}
\Delta (\mathbf{k,}i\omega _{n})=-T\sum_{\omega _{m},p}\frac{V_{pair}(%
\mathbf{k,p;}i\omega _{n},i\omega _{m})\Delta (\mathbf{p},i\omega _{m})}{%
(\omega _{m}+i\Sigma (\mathbf{p,}i\omega _{m}))^{2}+\xi _{\mathbf{p}%
}^{2}+\Delta ^{2}(\mathbf{p},i\omega _{m})}  \label{Eq:gap_eq}
\end{equation}%
assuming spin singlet pairing (for a review see \cite{Allen83}) . The pair
interaction is given as the sum of the screened Coulomb interaction and a
exchange contribution

\begin{equation}
V_{pair}(k\mathbf{,}p)=V_{C}(k-p)+V_{X}(k,p)
\end{equation}%
where $k=(\mathbf{k,}i\omega _{n}),$ etc., are momentum and Matsubara
frequency variables.

In doped SrTiO$_{3}$ the Coulomb interaction is screened by ionic and
electronic charges

\begin{equation}
V_{C}(q)=\frac{4\pi e^{2}}{q^{2}\varepsilon _{ion}(q)+4\pi e^{2}\chi _{el}(q)%
}
\end{equation}%
where $\varepsilon _{ion}(q)$ is the dielectric function of the undoped
system and $\chi _{el}(q)$ accounts for the screening effected by the
itinerant electronic charges. The ionic screening is dominated by the soft
transverse optical phonon mode ($\omega _{TO}$) and its longitudinal partner
($\omega _{LO}$, for a discussion see e.g. \cite{Vanderbilt94})

\begin{equation}
\varepsilon _{ion}(q)\approx \varepsilon _{\infty }\frac{\omega
_{n}^{2}+\omega _{LO}^{2}(\mathbf{q)}}{\omega _{n}^{2}+\omega _{TO}^{2}(%
\mathbf{q)}}
\end{equation}%
where $\varepsilon _{\infty }$\ is the optical dielectric constant. The
electronic response $\chi _{el}(q)$ is given by the bubble diagram and may
be approximated by

\begin{equation}
\chi _{el}(q)\approx \frac{n}{\frac{2}{3}\epsilon _{F}+\frac{\omega _{n}^{2}%
}{2\epsilon _{q}}+\frac{1}{2}\epsilon _{q}}
\end{equation}%
where $\epsilon _{q}=q^{2}/2m_{1}$ and $m_{1}$ is the carrier mass. At the
lowest densities $n$ for which superconductivity has been observed \cite%
{Behnia14} $n=5\times 10^{17}$cm$^{-3}$ , the Fermi energy is $\epsilon
_{F}\approx 14$K (we use energy units of Kelvin, and also take $\hbar =1$).
This should be compared with a typical phonon frequency $\omega _{D}=380$K .

As shown below, the exchange terms $V_{X}(k,p)$ of the interaction are
systematically small and may be neglected, except for a contribution
involving exchange of the soft phonon. The latter will be important near the
ferroelectric transition, but we will omit this contribution here.

Let us first discuss the behavior of the interaction function $V_{C}(\mathbf{%
q},i\omega _{l})$ for large momenta and frequencies. In the limit of large $%
q $ we have $V_{C}\propto q^{-2}$, which together with the fall off of the
Green's function provides convergence of the momentum integration. In the
limit of large frequency one finds $V_{C}=4\pi e^{2}/(q^{2}\varepsilon
_{\infty })$ , such that the frequency summation is not convergent unless
the high frequency behavior of the self energy $\Sigma (\mathbf{p,}i\omega
_{m})$ is taken into account. We estimated in our paper that the imaginary
part of $\Sigma (\mathbf{p,}\omega +i0)$ ($\Sigma (\mathbf{p,}i\omega _{m})$
analytically continued to the real frequency axis) grows with $\omega $ ,
exceeding $\omega $ beyond a frequency $\omega _{c}$. For $\omega >\omega
_{c}$ , $\Sigma (\mathbf{p,}i\omega _{m})>\omega _{m}$ and the Green's
function falls of faster than $\omega _{m}^{-2}$ thus ensuring convergence
of the $\omega _{m}$ summation in the gap equation. We therefore used $%
\omega _{c}$ as a density dependent frequency cutoff $\omega _{c}(n)$. We
found $\omega _{c}$ to vary with density in the range of $600$K to $2500$K,
which is much higher than the Fermi energy at low doping (see Fig.\ \ref%
{fig:coupl_vs_logn}) . A full-scale evaluation of the selfenergy is beyond
the scope of the present work.

\begin{figure}[tbp]
\includegraphics[width=1.2\columnwidth]{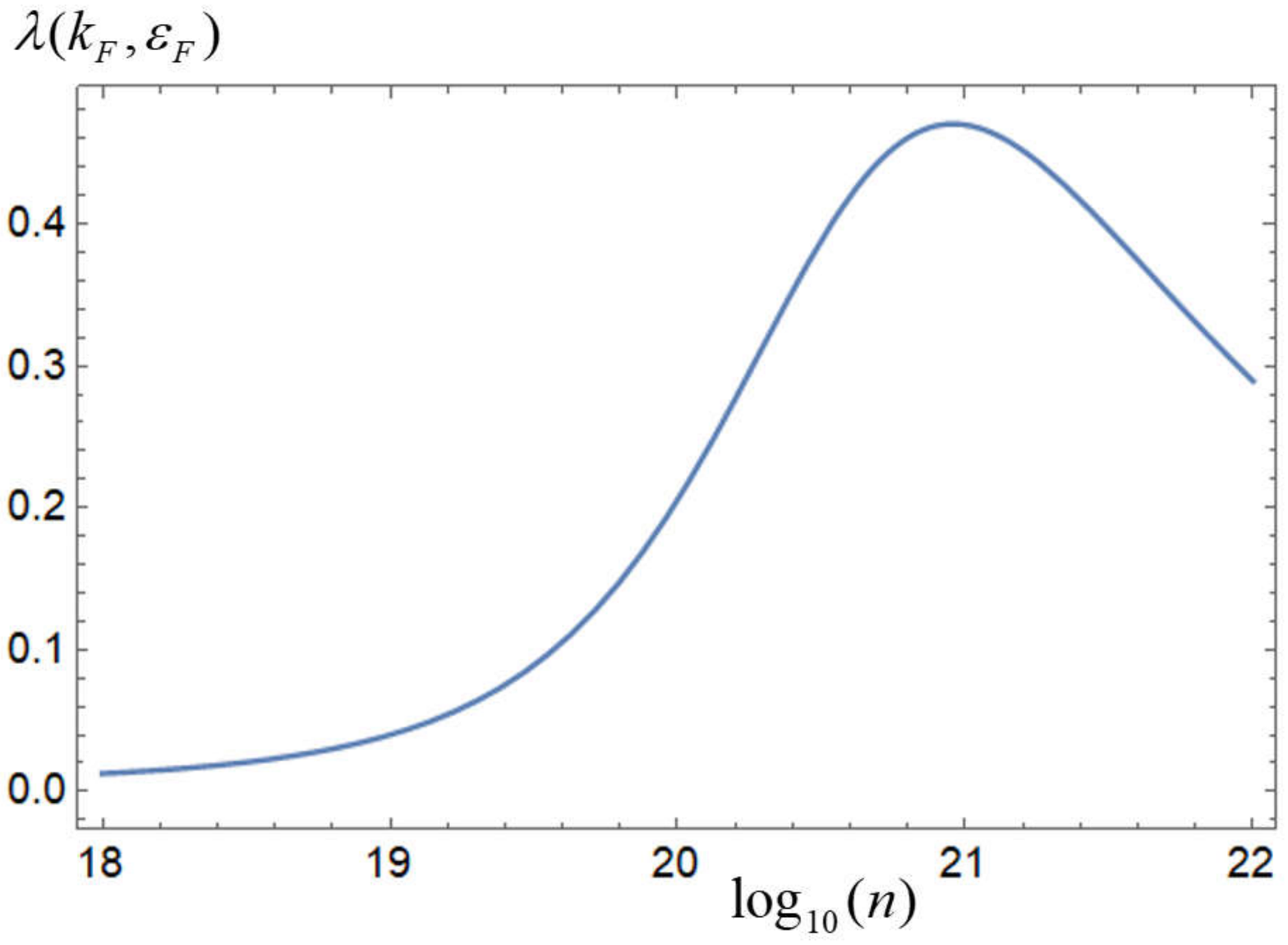}
\caption{Coupling function $\protect\lambda(k_{F},\protect\epsilon_{F})$
versus the logarithm of density}
\label{fig:coupl_vs_logn}
\end{figure}

\begin{figure}[tbp]
\includegraphics[width=1.2\columnwidth]{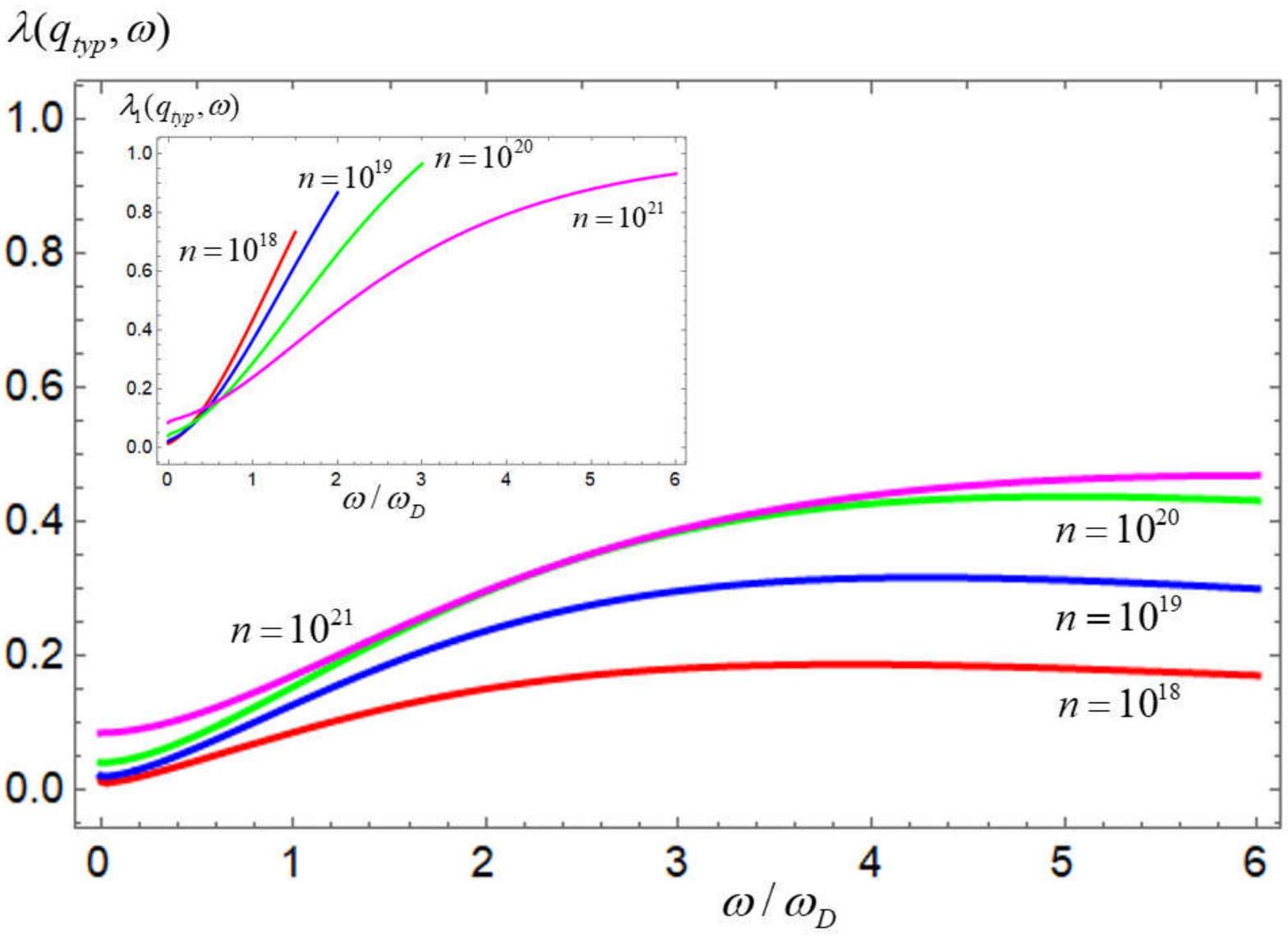}
\caption{Coupling function $\protect\lambda(q_{typ},\protect\omega)$ versus
frequency at various densities $n$. The inset shows coupling function $%
\protect\lambda_{1}(q_{typ},\protect\omega)$ versus frequency in the
relevant region $\protect\omega < \protect\omega_{c}$.}
\label{fig:coupl_vs_om}
\end{figure}

In order to justify the neglect of higher order correction terms to the gap
equation at such elevated frequencies we now show that the system indeed
remains in the weak coupling regime for frequencies up to $\omega _{c}(n)$.
We define a dimensionless coupling function $\lambda (q,\omega )=N(\epsilon
_{F})V_{C}(q\mathbf{,}i\omega )$ as a measure of the strength of the
interaction ($N(\epsilon _{F})=m_{1}k_{F}/\pi ^{2}$ is the density of states
at the Fermi energy). We adopt this definition of the coupling as the
relevant one, rather than the alternative definition $\lambda _{1}(q,\omega
)=N(\epsilon _{q})V_{C}(q\mathbf{,}i\omega )$, where $N(\epsilon
_{q})=m_{1}q/\pi ^{2}$, because $\lambda _{1}(q,\omega )$ is usually
multiplied by a factor $k_{F}/q$, and $\lambda _{1}(q,\omega
)(k_{F}/q)=\lambda (q,\omega )$. The factor $k_{F}/q$ arises in higher order
expressions such as vertex corrections, crossed interaction terms, and more,
from the angular integral of the vector $\mathbf{q}$ in the arguments of
single particle Green's functions, e.g. $G(\mathbf{k+q},i(\omega _{n}+\nu
_{m}))$. At higher electron densities the electron spectrum deviates from
the simple parabolic form, leading to a slower growth of the density of
states than assumed above, which also supports the choice of $\lambda
(q,\omega )$ as a preferable measure of the interaction strength. In Fig.\ %
\ref{fig:coupl_vs_logn} we first show the coupling function at the Fermi
energy $\lambda (k_{F},\epsilon _{F})$ versus electron density $n$ ($n$ in
units of cm$^{-3}$). At low density $\lambda (k_{F},\epsilon _{F})$ is seen
to be very much less than unity, getting larger at higher density, but
remaining less than unity in the whole density regime.\ One should note that
the cutoff frequency $\omega _{c}(n)$ approaches the Fermi energy for
densities $n\gtrsim 10^{20}$cm$^{-3}$ anyway, and then the issue of \ the
cutoff exceeding the Fermi energy is absent. The smallness of\ $\lambda
(k_{F},\epsilon _{F})$\ at low density suggests that $\lambda $ will remain
small at much higher frequencies and momenta.\ The coupling function
increases with frequency, but as shown in Fig.\ \ref{fig:coupl_vs_om} it
never exceeds unity, for frequencies $\omega <6\omega _{D}=2300$K. In Fig.\ %
\ref{fig:coupl_vs_om} the coupling is shown for the typical momentum at
frequency $\omega $, defined as $q_{typ}=k_{F}+\sqrt{2m_{1}\omega }$ . Also
shown is the dependence on density. For completeness we also show the
coupling function $\lambda _{1}(q_{typ},\omega )$\ in the inset of Fig.\ \ref%
{fig:coupl_vs_om}, which is still less than unity in the relevant
frequency range $\omega <\omega _{c}$.\ We may conclude from Figs.\ \ref%
{fig:coupl_vs_logn},\ \ref{fig:coupl_vs_om} that in the frequency range
up to $6\omega _{D}$ higher order contributions such as vertex corrections
and exchange interaction terms, which are of second or higher order in the
coupling $\lambda $ may be safely neglected. This conclusion holds provided
there is no additional instability in the system such as a charge or spin
density wave instability, in the neighborhood of which even a small
interaction may be critically enhanced. There is no experimental indication
of any other instability in addition to the ferroelectric instability. The
latter is expected to give rise to an enhanced pairing interaction mediated
by the soft transverse optical phonon close to the ferroelectric quantum
critical point.

\begin{figure}[tbp]
\includegraphics[width=1.2\columnwidth]{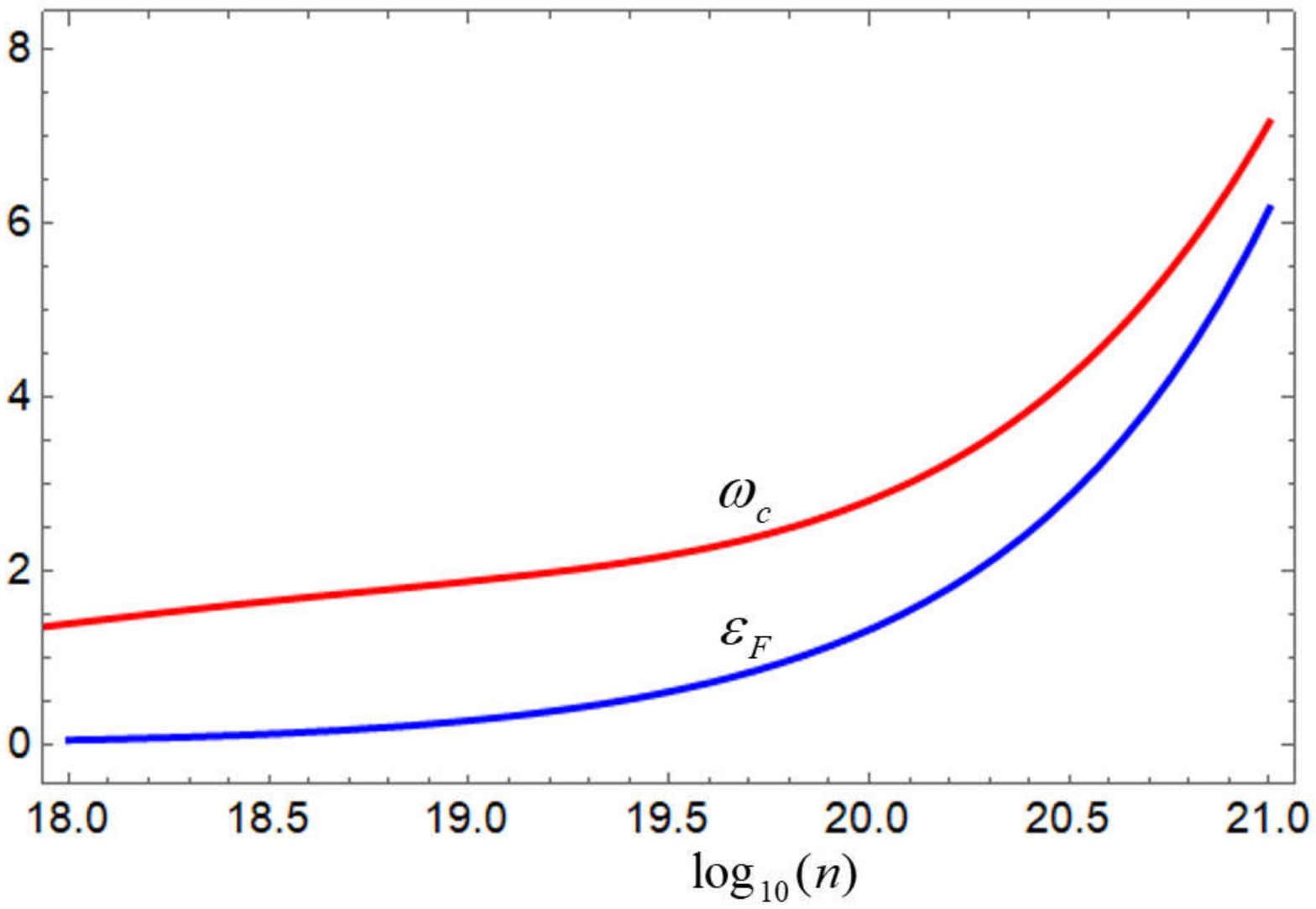}
\caption{Cutoff frequency $\protect\omega_{c}$ (red line) and Fermi energy
(blue line) in units of $\protect\omega_{D}$ versus the logarithm of density}
\label{fig:omc_vs_logn}
\end{figure}

In our paper we presented the results obtained for the combined pair
interaction $V_{C}+V_{X}$ with $V_{X}$ given by the TO-phonon mediated
exchange interaction, which was unfortunately not correctly derived. Now we
assume that $V_{X}$ is negligible if the system is not yet close to the
ferroelectric transition, i.e. in the absence of isotope doping or applied
pressure, but will be important close to the transition. We reevaluate the
transition temperature $T_{c}$ from Eq.\ \ref{Eq:gap_eq} above in the limit $%
\Delta \rightarrow 0$. For calculational convenience we again use a hard
frequency cutoff $\omega _{c}(n).$\ In our paper we distinguished three
different density regimes, for which we found different density power laws,
which we combined into an interpolation formula

\begin{equation}
\omega _{c}(n)=\frac{\omega _{D}}{(c_{1}(\frac{k_{F}}{q_{R}})^{-1/2}+c_{2}(%
\frac{k_{F}}{q_{R}})^{2})^{-1}+c_{3}(\frac{k_{F}}{q_{R}})^{-1}}
\end{equation}%
with parameters $c_{1}$, $c_{2}$, $c_{3}$ and where the Fermi wave number $%
k_{F}(n)=(3\pi ^{2}n)^{1/3}$and $q_{R}^{2}/2m=\omega _{D}$ have been used.
The prefactors of these power laws were estimated by order of magnitude in
Appendix A of \cite{Wolfle18}. In the numerical solution of the linearized
gap equation we assumed reasonable values for the prefactors such that the
resulting transition temperatures $T_{c}$ were in agreement with the data.
In Fig.\ \ref{fig:omc_vs_logn} the cutoff frequency is plotted as a function
of $n$, using the parameters $c_{j}$ specified below.  For comparison the
Fermi energy is shown as well. At low density $\omega_{c}$ is seen to be
orders of magnitude larger than the Fermi energy. At high density the cutoff
frequency approaches the Fermi energy from above.

In Matsubara space $V_{C}(\mathbf{q},i\omega _{l})$ is a positive definite
function, which implies that the eigenfunctions $\Delta (\mathbf{k,}i\omega
_{n})$ of the linearized gap equation must necessarily have zero's in
frequency space\cite{Bogoliubov58,Morel-Anderson62}. We find an even
frequency eigenfunction $\Delta (\mathbf{k,}i\omega _{n})$ featuring two
zero's at $\pm \omega _{n_{0}}$.

\begin{figure}[tbp]
\includegraphics[width=1.2\columnwidth]{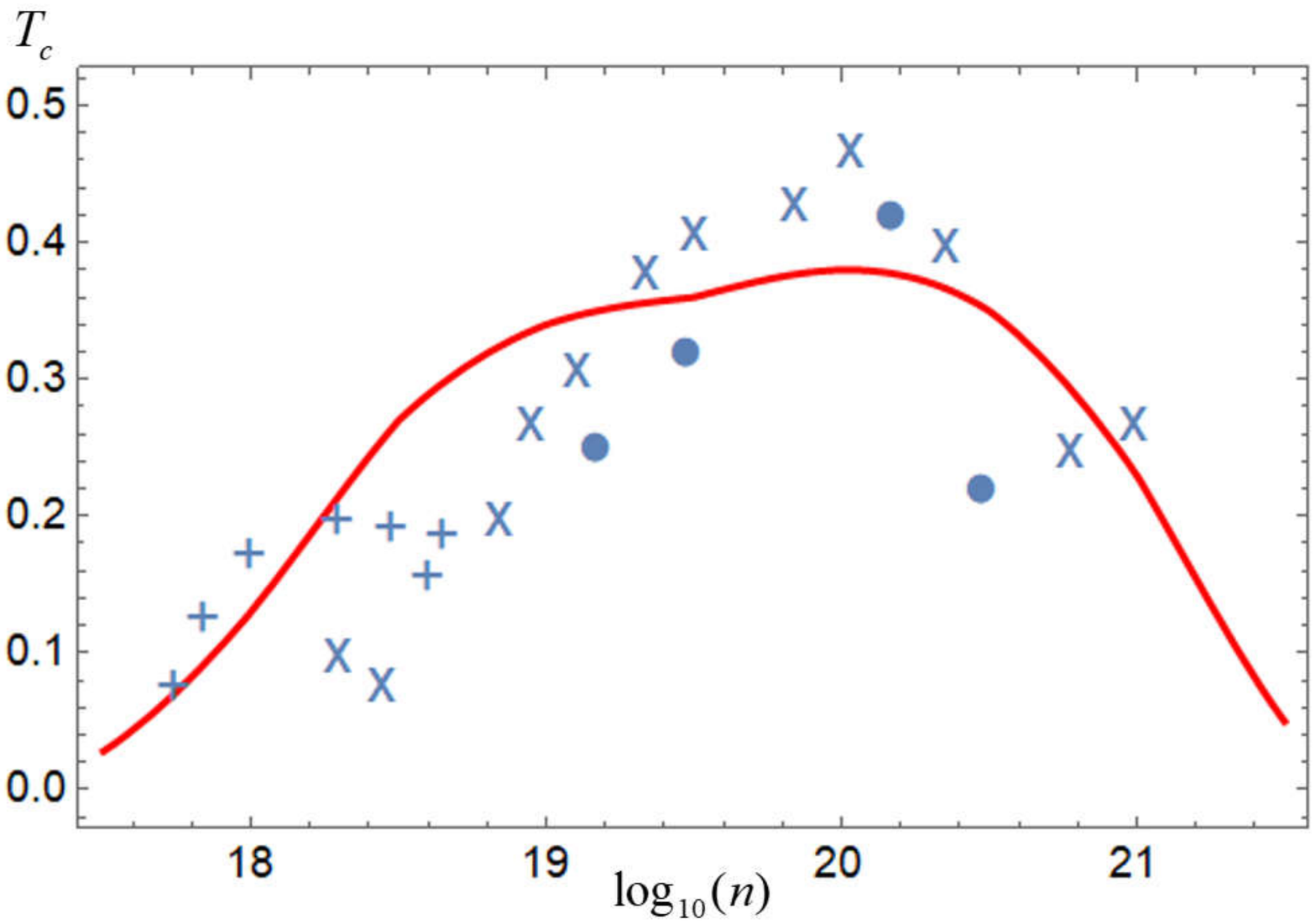}
\caption{Transition temperature Tc in Kelvin versus logarithm of electron
density in cm$^{-3}$. Theory: solid line; Experiment: crosses (Schooley et
al., 1964); filled circles: Nb-doped; + symbols: O-reduced (Lin et al.,
2014) }
\label{fig:Tc_vs_logn}
\end{figure}

In Fig.\ \ref{fig:Tc_vs_logn} the result for $T_{c}$ versus $log_{10}n$ is
shown for the choice of parameters $(c_{1},c_{2},c_{3})=(0.49,1.55,0.118)$
as compared to the values $(1.1,0.6,0.036)$ used in our earlier evaluation 
\cite{Wolfle18} for the combined pair interaction in which the attractive TO
phonon mediated interaction was taken sufficiently strong so that the
eigenfunction did not have zeros. It should be noted that the two components
of the pair interaction in that calculation \cite{Wolfle18} partially
compensate each other, leading to a distinctly different frequency
dependence, which explains the change in the $c_{j},$ $j=1,2,3$ . A more
quantitative evaluation of the parameters $c_{j}$, for the two cases of pair
interaction, $V_{C}$, and $V_{C}+V_{X}$ would lead to somewhat different
values.\ Also shown in Fig.\ \ref{fig:Tc_vs_logn} are experimental data, which
are reasonably well accounted for by our theory.

\section{The weakly attractive Fermi gas}

In order to clarify the question of the energy cutoff in the gap equation
further it is useful to consider a system of weakly interacting fermions
allowing for controlled approximations. We consider a Fermi gas with
quasiparticle energy $\xi _{\mathbf{k}}=(k^{2}-k_{F}^{2})/2m$ and weak
attractive interaction of the form

\begin{equation}
V_{pair}(\mathbf{k},\mathbf{k}^{\prime })=\left\{ 
\begin{array}{c}
-V_{0},\text{ \ \ \ \ }|\xi _{\mathbf{k}}|,|\xi _{\mathbf{k}^{\prime
}}|<p_{c}^{2}/2m \\ 
0,\text{ \ \ \ \ \ \ else}%
\end{array}%
\right.
\end{equation}%
We now assume that $p_{c}\gg k_{F}$ and that the dimensionless coupling $%
\lambda (\epsilon )=V_{0}N(\epsilon )\ll 1$ , for $\epsilon <\epsilon
_{c}=p_{c}^{2}/2m$, where $N(\epsilon )=mk/\pi ^{2}$ and $\epsilon =k^{2}/2m$%
. In this case higher order terms in perturbation theory contributing to
both, the irreducible particle-particle vertex (the full pair interaction)
and the self energy are negligible. The linearized gap equation takes the
form

\begin{equation}
\Delta =-\sum_{\mathbf{p}}V_{pair}(\mathbf{k},\mathbf{p})\frac{\tanh \frac{%
\xi _{\mathbf{p}}}{2T_{c}}}{2\xi _{\mathbf{p}}}\Delta
\end{equation}%
The transition temperature follows as \cite{Gorkov61}

\begin{equation}
T_{c}\approx \epsilon _{F}\exp [-\frac{1}{N_{0}|\Gamma _{0}|}]
\label{Tc_weak}
\end{equation}%
where $\Gamma _{0}$ may be interpreted as scattering amplitude. For the
above model one finds

\begin{equation}
\frac{1}{N_{0}|\Gamma _{0}|} =(\frac{1}{N_{0}V_{0}}-\frac{p_{c}}{k_{F}}%
)(1+O(\lambda(\epsilon_{c}))
\end{equation}%
where $N_{0}=mk_{F}/2\pi ^{2}$ is the density of states (of one spin
component) at the Fermi level. In the usual limit of attraction only within
a narrow energy shell about the Fermi level , $p_{c}\ll k_{F}$ , the weak
coupling result is recovered, $T_{c}\approx \epsilon _{c}\exp [-\frac{1}{%
N_{0}V_{0}}]$. In the opposite case, $p_{c}\gg k_{F}$, which is the
situation realized in weakly doped SrTiO$_{3}$, the transition temperature
may be enhanced by orders of magnitude due to the strong renormalization of
the scattering amplitude $\Gamma _{0}$ by virtual processes involving states
in the high energy region $\epsilon _{F}<\xi _{\mathbf{p}}<\epsilon _{c}$.

\section{Conclusion}

The observed superconductivity of lightly doped STO challenges the
"conventional wisdom" accumulated over 60 years of theoretical effort to
understand many different manifestations of this phenomenon. Fortunately,
the situation here is much simpler than that posed by the enigmatic strong
coupling superconductors, in particular the cuprates, in that the charge
carriers in STO are in the weak coupling limit. The extremely strong ionic
screening reduces the strength of the (dimensionless) dynamic Coulomb
interaction to about $0.01$ to $0.2$, depending on density and energy, up to
the LO phonon energy. As a consequence, the contribution of high energy ($%
\epsilon >\epsilon _{F}$) virtual processes to pairing is not cut off by
higher order contributions such as vertex corrections, but is found to
enhance the critical temperature at the lowest densities by orders of
magnitude.This physics has been known by the pioneers of superconductivity,
e.g. Gorkov and Melik-Barkhudarov, who stated that for a Fermi gas with weak
attraction the bare interaction has to be replaced by the scattering
amplitude, which may be much larger than the bare interaction owing to
scattering into high energy intermediate states. The question is then what
determines the frequency cutoff in the gap equation. Here we propose that
the growth of the electron self\ energy $\Sigma (\omega )$ with energy
provides a cutoff at $\omega _{c}=\Sigma (\omega _{c})$. The strong fall off
of the Green's \ function with energy beyond that point ensures convergence
of the frequency summation in the gap equation. The ultimate energy limit $%
\omega _{x}$ beyond which the weak coupling treatment would no longer be
valid is given by the energy at which the dimensionless coupling $%
\lambda(\omega _{x})\approx 1$. As shown above (see Figs.\ \ref%
{fig:coupl_vs_logn},\ \ref{fig:coupl_vs_om_v2}), $\omega _{c}\ll \omega _{x}$
, implying that $\omega _{x}$ is not relevant.\ The effect of the cutoff $%
\omega_{x}$ has been discussed in the context of the plasmon exchange
mechanism for conventional metals \cite{Grabowski84}. A different lesson to
be learned from the pioneers, here Bogoliubov and Anderson, is that in the
case of the pair interaction given by the screened Coulomb interaction,
which is a positive definite function (in Matsubara frequency space) the gap
function must change sign as a function of frequency (leading for
even-frequency pairing to a kind of "d-wave" pairing in frequency space).
The observed "dome" in $T_{c}$ versus $\log n$ can be fully accounted for by
the limiting effect of strong electronic screening on the high density side
and the cutoff energy $\omega _{c}\propto k_{F}^{1/2}$ derived from
electron-phonon scattering at low density. As for the observed strong
isotope effect, which may be expected to arise from a pair interaction
contribution mediated by the soft TO phonon mode, we have to withdraw our
earlier proposal of a deformation potential electron-phonon coupling. We do,
however, take the present observations of the isotope effect at face value,
noting that in the meantime new data on the effect of strain on $T_{c}$ also
suggest a strong coupling of carriers to the soft mode\cite{Sochnikov18}.
Work on deriving a sufficiently strong alternative e-ph coupling is in
progress.

\section{Acknowledgements}

PW acknowledges support by a Distinguished Senior Fellowship of Karlsruhe
Institute of Technology. AVB is supported by ERC Synergy HERO (810451) and
KAW 2018-0104.


\begin{thebibliography}{99}

\bibitem{Ruhman19} J. Ruhman and P. Lee, Comment (2019).

\bibitem{Wolfle18} P. W\"{o}lfle, and A. V. Balatsky, Phys. Rev. B \textbf{98%
}, 104505 (2018).

\bibitem{Ruhman16} J. Ruhman, and P. A. Lee, Phys. Rev. B \textbf{94},
224515 (2016).

\bibitem{Behnia13} X. Lin, Z. Zhu, B. Fauque%
\'{}%
, and K. Behnia, Phys. Rev. X \textbf{3}, 021002 (2013).

\bibitem{Behnia14} X. Lin, G. Bridoux, A. Gourgout, G. Seyfarth, S. Kramer,
M. Nardone, B. Fauque%
\'{}
and K. Behnia, Phys. Rev. Lett. \textbf{112}, 207002 (2014).

\bibitem{Schooley64} J. F. Schooley, W. R. Hosler, and M. L. Cohen, Phys.
Rev. Lett. \textbf{12}, 474 (1964).

\bibitem{vdMarel16} A. Stucky, G. W. Scheerer, Z. Ren, D. Jaccard, J.-M.
Poumirol, C. Barreteau, E. Giannini and D. van der Marel, Nature Scientific
Reports \textbf{6, }37582 (2016).

\bibitem{Sochnikov18} C. Herrera et.al, Strain-engineered interaction of
quantum polar and superconducting phases, arXiv:1808.03739 (2018).

\bibitem{Eliashberg60} G. M. Eliashberg, Sov. Phys. Sol. JETP \textbf{11},
696 (1960).

\bibitem{Allen83} P. B. Allen, and B. Mitrovic, Solid State Physics, Vol. 37, 
pp. 1-92 (1983).

\bibitem{Vanderbilt94} W. Zhong, R.D. King-Smith, and David Vanderbilt,
Phys. Rev. Lett. \textbf{72}, 3618 (1994).

\bibitem{Bogoliubov58} N. N. Bogoliubov, V. V. Tolmachev, and D. V. Shirkov, 
\textit{A New Method in the Theory of Superconductivity} (1958)

(translation: Consultants Bureau, Inc. , New York, 1959).

\bibitem{Morel-Anderson62} P. Morel, and P.W. Anderson, Phys. Rev. \textbf{%
125}, 1263 (1962).

\bibitem{Gorkov61} L. P. Gor'kov, and T. K. Melik-Barkhudarov, Soviet
Physics JETP \textbf{13}, 1018 (1961).

\bibitem{Grabowski84} M. Grabowski and L. J. Sham, Phys. Rev. B \textbf{29},
6132 (1984).
\end{thebibliography}
\end{document}